\documentclass[aps,prx,twocolumn,showpacs,superscriptaddress]{revtex4-1}

\usepackage{graphicx}
\usepackage{dcolumn}
\usepackage{bm}
\usepackage{amsmath} 
\usepackage[dvips]{color}

\newcommand{\beps}{\boldsymbol\epsilon}
\begin{document}

\title{Differences in the quasiparticle dynamics for one-band and
  three-band cuprate  models }  
\author{Hadi Ebrahimnejad} \affiliation{Department of Physics and
  Astronomy, University of British Columbia, Vancouver, BC, Canada,
  V6T 1Z1}

\author{George A. Sawatzky} \affiliation{Department of Physics and Astronomy,
  University of British Columbia, Vancouver, BC, Canada, V6T 1Z1}
\affiliation{Quantum Matter Institute, University of British Columbia,
  Vancouver, BC, Canada, V6T 1Z4}

\author{Mona Berciu} \affiliation{Department of Physics and Astronomy,
  University of British Columbia, Vancouver, BC, Canada, V6T 1Z1}
\affiliation{Quantum Matter Institute, University of British Columbia,
  Vancouver, BC, Canada, V6T 1Z4}

\begin{abstract}
We study the quasiparticles of the one-band $t$-$J$ and
$t$-$t'$-$t''$-$J$ models using a variational approximation that
includes spin fluctuations in the vicinity of the hole. We explain why
the spin fluctuations and the longer range hopping have complementary
contributions to the quasiparticle dynamics, and thus why both are
essential to obtain a dispersion in agreement with that measured
experimentally. This is very different from the three-band Emery model
in the strongly-correlated limit, where the same variational
approximation shows that spin fluctuations have a minor effect on the
quasiparticle dynamics. This difference proves that these one-band and
three-band models describe qualitatively different quasiparticles, and
therefore they they cannot both be suitable to describe the physics of
underdoped cuprates.
\end{abstract}

 \maketitle

\section{Introduction}

Nearly three decades after their discovery \cite{Bend}, the
high-temperature cuprate superconductors have so far eluded a
comprehensive explanation. These layered materials contain CuO$_2$
layers, which are antiferromagnetic insulators in the undoped
limit and become superconducting upon doping
\cite{RMP_reviews}. The hole-doped side shows a more robust
superconductivity, extending to higher temperatures and over a wider
range of dopings. The first step towards deciphering the mechanism of
superconductivity in these compounds is a proper description of the
motion of these holes in the CuO$_2$ layer -- this has become one of
the most studied problems in condensed matter theory
\cite{RMP_reviews,Kane89}. Its solution should elucidate the nature of
the quasiparticles that eventually bind together into the pairs that
facilitate the superconducting state.

Despite significant effort, even what is the minimal model that
correctly describes this low-energy quasiparticle is still not
clear. There is general agreement that the parent compounds are
charge-transfer insulators \cite{Zaanen}, and wide consensus that most
of their low-energy physics is revealed by studies of a single CuO$_2$
layer, modeled in terms of Cu $3d_{x^2-y^2}$ and O $2p$ orbitals.
Because only ligand $2p$ orbitals hybridize with the $3d_{x^2-y^2}$
orbitals, it is customary to ignore the other O $2p$ orbitals; this
leads to the well-known three-band Emery model \cite{Emery}.

However, the Emery model is perceived as too complicated to study so it
 is often further simplified to a one-band $t$-$J$ model that
 describes the dynamics of a Zhang-Rice singlet (ZRS)
 \cite{Zhang,George}.  We known that the $t$-$J$ model with only
 nearest-neighbor (nn) hopping $t$ is certainly not the correct model
 because it predicts a nearly flat quasiparticle energy along
 $(0,\pi)-(\pi,0)$, unlike the substantial dispersion found
 experimentally \cite{Wells,Leung95,Damascelli}. However, its
 extension with longer range hopping, the $t$-$t'$-$t''$-$J$ model,
 was shown to reproduce the correct dispersion \cite{Leung97} for
 values of the 2$^{\rm nd}$ and 3$^{\rm rd}$ nn hoppings
 in rough agreement with those estimated from density functional theory
 \cite{ok} and cluster calculations \cite{eskes}. Taking this
 agreement as proof that this is the correct model, both the $t$-$J$
 model and its parent, the Hubbard model \cite{amol}, have been
 studied very extensively in the context of cuprate physics.

Here we show that the quasiparticle of the $t$-$t'$-$t''$-$J$ model is
{\em qualitatively} different from that of the $U_{dd} \rightarrow
\infty$ limit of the three-band Emery model \cite{comment}. While both
models predict a dispersion in quantitative agreement with that
measured experimentally (for suitable values of the parameters), the
factors controlling the quasiparticle dynamics are very different. It
has long been known that both the longer-range hopping and the spin
fluctuations play a key role in the dynamics of the quasiparticle of
the $t$-$t'$-$t''$-$J$ model. Here we use a non-perturbative
variational method, which agrees well with available exact
diagonalization (ED) results, to show that spin-fluctuations and
longer-range hopping control the quasiparticle dispersion in different
parts of the Brillouin zone, and to explain why.  In contrast, using
the same variational approach, it was recently argued that spin
fluctuations play no role in the dispersion of the quasiparticle of
the $U_{dd} \rightarrow \infty$ limit of the Emery model
\cite{Hadi}. This claim is supported  by additional
results we present here.

This major difference in the role played by spin fluctuations in
determining the quasiparticle dynamics shows that these models do
not describe the same physics. This suggests that the
$t$-$t'$-$t''$-$J$ model is not suitable for the study of cuprates in
the hole-doped regime, although it and related one-band models may
be valid in the electron-doped regime. As we argue below, it may
be possible to ``fix'' one-band models by addition of other terms,
although we do not expect this to be a fruitful enterprise. Instead,
we believe that what is needed is a concerted effort to understand the
predictions of the Emery model. Our results in Ref. \cite{Hadi} and
here that spin fluctuations of the AFM background do not play a key
role in the quasiparticle dynamics of this three-band model, contrary to what was believed to
be the case based on results from one-band models, should simplify
this task.

The article is organized as follows. In Section II we review the
three-band Emery model and briefly discuss the emergence of the
one-band and simplified three-band models in the asymptotic limit of
strong correlations on the Cu sites. Section III describes the
variational method, which consists in keeping a limited number of
allowed magnon configurations in the quasiparticle cloud. Section IV
presents our results for both one- and three-band models, and their
interpretation. Finally, Section V contains a summary and a detailed discussion
of the implications of these results.

\section{Models}

\begin{figure}[t]
\centering
\includegraphics[width=\columnwidth]{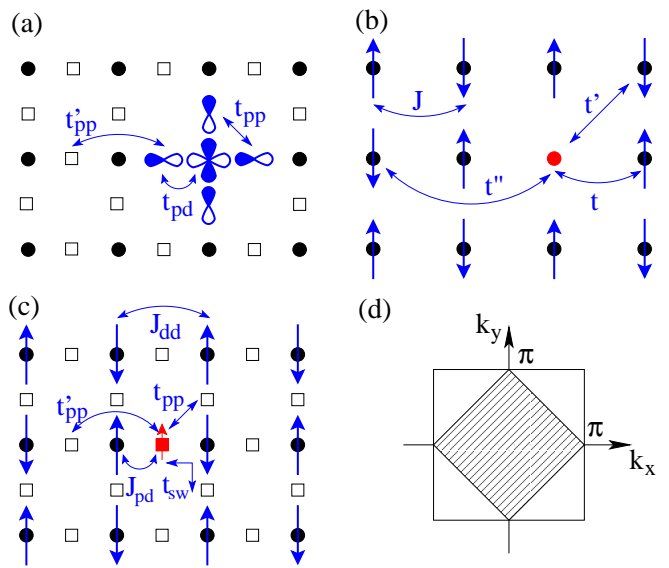}
\caption{(color online) (a) Sketch of the CuO$_2$ layer, with O are
  marked by squares  and Cu marked by circles. The relevant orbitals
  are drawn at a few sites. The arrows indicate the hopping terms
  included in the three-band Emery model; (b) Sketch of the one-band
  $t$-$t'$-$t''$-$J$ model. Cu sites host spin degrees of freedom,
  except at sites where a Zhang-Rice singlet is centered (red
  circle). The arrows indicate the various terms in this Hamiltonian;
(c) Sketch of the $U_{dd}\rightarrow \infty$ limit of the Emery
  model. Cu sites  host spin degrees of freedom but the doped holes
  (red filled square) move on the O lattice. The arrows indicate the
  various terms in this Hamiltonian; (d) 
The full Brillouin zone of the CuO$_2$ lattice (the outer square)
which encloses the magnetic Brillouin zone (shaded). }   
\label{fig1}
\end{figure}

A widely accepted starting point for the description of a CuO$_2$
layer is the three-band Emery model \cite{Emery}:
\begin{multline}
\label{Emr}
\mathcal{H} = T_{pp} + T_{pd} + \Delta_{pd}\sum_{i \in {\rm O},
  \sigma}n_{i, \sigma} \\+ U_{pp}\sum_{i \in {\rm O}}n_{i,
  \downarrow}n_{i, \uparrow} + U_{dd}\sum_{i \in {\rm Cu}}n_{i,
  \downarrow}n_{i, \uparrow}.
\end{multline}
The sets ${\rm O}$ and ${\rm Cu}$  contain the ligand O $2p$ 
and the Cu $3d_{x^2-y^2}$ orbitals respectively, sketched in
Fig. \ref{fig1}(a). For ${i \in {\rm O}}$, $n_{i,
  \sigma}=p^{\dagger}_{i,\sigma}p_{i, \sigma}$ is the number of
spin-$\sigma$ holes in that $2p$ orbital. Similar notation is used for
the $3d$ orbitals, their hole creation operators being
$d^\dagger_{i,\sigma}$, $i\in {\rm Cu}$.

$T_{pp}$ is the kinetic energy of the holes moving on the O
sublattice, described by a Hamiltonian with first ($t_{pp}$) and
second ($t'_{pp}$) nearest-neighbour (nn) hopping:
\begin{multline}
T_{pp} = t_{pp}\sum_{i\in {\rm O},{\boldsymbol\delta},\sigma}
r_{\boldsymbol\delta}p^{\dagger}_{i,{\sigma}}p^{}_{i+{\boldsymbol\delta},\sigma}\\ -
t'_{pp}\sum_{i \in {\rm O},\sigma} p^{\dagger}_{i,\sigma}(p^{}_{i -
  {\boldsymbol\epsilon},\sigma}+p^{}_{i +
  {\boldsymbol\epsilon},\sigma}).
 \label{Tpp}
\end{multline}
The lattice constant is set to $a=1$. The vectors
${\boldsymbol\delta}= \pm(0.5, 0.5), \pm(0.5,-0.5)$ are the distances
between any O and its four nn O sites, and $r_{{\boldsymbol\delta}}
= \pm  1$ sets the sign of each nn $pp$ hopping integral in accordance
with the overlap of the $2p$ orbitals involved, see
Fig. \ref{fig1}(a).  Next nn hopping is included only between O $2p$
orbitals pointing toward a common bridging Cu, separated by
${\boldsymbol\varepsilon}= (1,0)$ or $(0,1)$; hybridization with
the $4s$ orbital of the bridging Cu further boosts the value of this hopping
integral.

$T_{pd}$ is the kinetic energy of holes moving between neighbour Cu
and O orbitals:
\begin{equation}
T_{pd} = t_{pd}\sum_{i \in {\rm Cu},\bf u,\sigma} r_{\bf
  u}d^{\dagger}_{i,{\sigma}}p^{}_{i+ \bf u,\sigma} + h.c.,
\end{equation}
where ${\bf u} = (\pm 0.5,0), (0, \pm 0.5)$ are the distances between
a Cu and its four nn O sites, and $r_{\bf u}$ are the signs of the
overlaps of the corresponding orbitals. It is this term that provides
the main justification for ignoring the other sets of $2p$ orbitals,
because symmetry forbids hopping of Cu holes from $3d_{x^2-y^2}$
orbitals into the non-ligand O orbitals. We further discuss this
assumption below.

$\Delta_{pd}$ is the charge transfer energy which ensures that in the
parent compound the O $2p$ orbitals are fully occupied ({\em i.e.}
contain no holes). Finally, $U_{pp}$ and $U_{dd}$ are the Hubbard
repulsion in the $2p$ and $3d$ orbitals, respectively. Longer range Coulomb
interaction between holes on O and Cu can also be added, but for the
single doped-hole problem analyzed here, it leads to a trivial
energy shift.

Strong correlations due to the large $U_{dd}$,
combined with the big Hilbert space with its three-orbitals basis,
and the need for a solution for a hole concentration equal (in the
parent compound) or  larger (in the hole doped case) than one
per Cu, make this problem very difficult to solve. While
progress is been made with a variety of techniques  \cite{3band} (which however
have various restrictions, such as rather high-temperatures and/or
small clusters for quantum Monte Carlo methods, and/or additional
approximations, such as setting $t_{pp}=0$ for convenience),  it is
far more customary to further simplify this model before attempting a solution.

A reasonable way forward is to use the limit $U_{dd} \rightarrow
\infty$ to forbid double occupancy of the $3d$ orbitals. As a result,
in the undoped ground-state there is one hole per Cu. Virtual hopping
processes lead to antiferromagnetic (AFM) superexchange between the resulting
 spin degrees of freedom, so that the parent compound is a
Mott insulator with long-range AFM order \cite{comm0}.

Upon doping, holes enter the O band  and the issue
is how to accurately describe their dynamics as they
interact with the spins at the Cu sites. Here we compare two such
descriptions for {\em the single doped hole case}.

\subsection{One-band  models}

In their seminal work \cite{Zhang}, Zhang and Rice argued that the
doped hole occupies the linear combination of the four $O$ 2p ligand
orbitals surrounding a central Cu, that has the same ${x^2-y^2}$
symmetry like the Cu $3d$ orbital hosting the spin. Furthermore,
exchange locks the two holes in a low-energy Zhang-Rice singlet
(ZRS). They also argued that the dynamics of this composite object,
which combines charge and spin degrees of freedom, is well captured by
the one-band model:
\begin{equation}
\label{Hamm}
{\cal{H}} = {\cal P}\left[\hat{T} + \hat{T'}+ \hat{T}''\right]{\cal
  P}+ \mathcal{H}_\mathrm{AFM}.
\end{equation} 
The first term describes the hopping of the ZRS (marked in
Fig. \ref{fig1}(b) by the ``missing spin'' locked in the ZRS) on the
square lattice of Cu sites that hosts it. Originally only nn hopping
$T$ was included: $\hat{T} = t \sum_{\langle i, j\rangle}^{}
d^\dagger_{i,\sigma} d_{j,\sigma} + h.c.$, with $i,j \in {\rm
  Cu}$. The projector ${\cal P}$ removes doubly occupied states,
therefore this term allows only Cu spins neighboring the ZRS to
exchange their location with the ZRS. This mimics the more complex
reality of the doped hole moving on the O sublattice and forming ZRS
with different Cu spins.

Although in Ref. \cite{Zhang} it was argued that only nn ZRS hoping  is
important, longer-range 2nd ($\hat{T}'$)
and 3rd ($\hat{T}''$) nn hopping was later added the model on a
rather ad-hoc basis. As discussed below, this is needed in order to
find a quasiparticle dispersion similar to that measured
experimentally. These terms are defined similarly to
$\hat{T}$ with hopping integrals $t'$ and $t''$, respectively. For
cuprates, $t'/t\sim-0.3, t''/t\sim 0.2$ are considered to be representative
values \cite{ts0, ts}, in agreement with various estimates \cite{ok,eskes}. In the 
following, we refer to this as the $t$-$t'$-$t''$-$J$ model, whereas if
$t'=t''=0$ we call it the $t$-$J$ model.

The term $\mathcal{H}_\mathrm{AFM} = J\sum_{\langle i,j\rangle}{\bf
  S}_i\cdot {\bf S}_{j}$ describes the nn AFM superexchange between
the other Cu spins ${\bf S}_i$, with $J/t\sim 0.3$ for cuprates
\cite{ts}. It leads to AFM order in the undoped system \cite{comm0},
and also controls the energy of the cloud of magnons that are created
in the vicinity of the ZRS, as it moves through the magnetic
background.

The $t$-$J$ model also emerges as the $U\rightarrow \infty$
limit of the Hubbard model \cite{amol}, but with additional terms of
order $J$. One of them, $-J/4 \sum_{\langle i, 
  j\rangle}^{}n_i n_j$, gives trivial energy shifts for both the
undoped and the single-hole doped cases of interest to us in this work, so its
presence can be safely ignored in this context. More interesting is
the so-called three-site term ${\cal P}\hat{T}_{3s} {\cal P}$ 
\cite{3site}, where 
\begin{multline}
\label{3site}
\hat{T}_{3s} = \frac{J}{4}\sum_{i \in {\rm Cu}, \sigma}\sum_{\boldsymbol
  \epsilon\neq\boldsymbol \epsilon'}(d^{\dagger}_{i + \boldsymbol \epsilon',
  \sigma}n_{i,-\sigma}d_{i + \boldsymbol \epsilon, \sigma} \\
 -d^{\dagger}_{i + \boldsymbol \epsilon',
   \sigma}d^{\dagger}_{i,-\sigma}d_{i, \sigma}d_{i + \boldsymbol
   \epsilon,-\sigma}) 
\end{multline}
describes ZRS hopping through an intermediate Cu site and permits spin
swapping with the spin at this intermediate site. As shown below, this
term influences the quasiparticle dispersion but it is not clear that
it should be included in the one-band model, because the original
Hamiltonian is the Emery, not the Hubbard, model.

In fact, a perturbational derivation of the low-energy Hamiltonian
obtained by projecting the three-band model onto ZRS states reveals a
much more complicated Hamiltonian than the $t$-$t'$-$t''$-$J$ model,
with many other terms \cite{ALigia}. We are not aware of a systematic
study of their impacts, but their presence underlies one important
issue with this approach: the hoped-for simplification due to the
significant decrease in the size of the Hilbert space comes at the
expense of a Hamiltonian whose full expression \cite{ALigia} is very
complicated. Using instead simpler versions like the
$t$-$t'$-$t''$-$J$ model may result in qualitatively different physics
than that of the full one-band model. Here we argue that
this is indeed the case.

\subsection{Simplified three-band model}

An alternative is to begin at the same starting point, {\em i.e.}
the limit $U_{dd} \rightarrow \infty$ resulting in spin degrees of
freedom at the Cu sites. However, the O sublattice on which the doped
hole moves is kept in the model, not projected out like in the
one-band approach, see Fig. \ref{fig1}(c). This leads to a bigger
Hilbert space than for
one-band models (yet smaller than for the Emery model) but because
spin and charge degrees of freedom are no longer lumped together, the
resulting low-energy Hamiltonian is simpler and makes it easier to
understand its physics.

The effective model for a layer with a single doped
hole, which for convenience we continue to call ''the three-band model''
although it is its $U_{dd}\rightarrow \infty$  approximation, 
was derived in Ref. \cite{Bayo} and reads: 
\begin{equation}
\label{eff}
\mathcal{H}_\mathrm{eff} = \mathcal{H}_\mathrm{AFM} +
\mathcal{H}_{J_{pd}} + T_{pp} +T_\mathrm{swap}. 
\end{equation}
The meaning of its terms is as follows:
$$\mathcal{H}_\mathrm{AFM} = J_{dd}\sum_{\langle i, j \rangle'} {\bf
  S}_{i}\cdot{\bf S}_{j}$$ is again the nn AFM superexchange between
the Cu spins ${\bf S}_{i}$, so $J_{dd}\equiv J$ of
the one-band models. The main difference is indicated by the presence of the
prime, which  reflects the absence of coupling for
the pair that has the doping hole on their bridging O. The next term, 
$$\mathcal{H}_{J_{pd}}= J_{pd} \sum_{i\in O, {\bf u}} {\bf
  s}_{i}\cdot{\bf S}_{i+{\bf u}}$$ 
is the exchange
of the hole's spin  ${\bf
  s}_{i} = {1\over 2} \sum_{\alpha,\beta}^{} p^\dagger_{i\alpha} {\boldsymbol
  \sigma}_{\alpha\beta} p_{i\beta}$ with its two nn Cu
spins. It arises from virtual hopping of a hole between a Cu and the O
hosting the doped hole.

Like in Eq. (\ref{Emr}), $T_{pp}$ is the kinetic energy of the doping
hole as it moves on the O sublattice. It is supplemented by
$T_\mathrm{swap}$ which describes effective hopping mediated by virtual
processes where a Cu hole hops onto an
empty O orbital, followed by the doping hole filling the now empty Cu
state \cite{Bayo}. This results in effective nn or next nn hopping of the
doped hole, with a swapping of its spin  with that of
the Cu involved in the process. The explicit form of this term is:
$$
T_\mathrm{swap} = -t_{sw}\sum_{i \in {\rm Cu}, {\bf u \ne  u'}}\sum_{
  \sigma, \sigma'} s_{\bf u-u'}p^{\dagger}_{i+ \bf
  u,{\sigma}}p^{}_{i+\bf u',\sigma'}|i_{\sigma'}\rangle\langle i_\sigma|,
 \label{Ts}
$$
reflecting the change of the Cu spin located at ${\bf R}_i$
from $\sigma$ to $\sigma'$ as the hole changes its spin from
$\sigma'$ to $\sigma$ while moving to another O. The sign
$s_{\boldsymbol\eta}=\pm 1$ is due to the overlaps of the 2$p$ and
3$d$ orbitals involved in the process.

For typical values  
of the parameters of the Emery model   \cite{RMP_reviews} and using 
$J_{dd}$ ($\sim 0.15$ eV) as the unit of energy, the dimensionless
values of the other parameters are
$t_{pp}\sim 4.1$, $t'_{pp}\sim 0.6t_{pp}$, $t_{sw}\sim 3.0 $ and
$J_{pd}\sim 2.8$. We use these values in the following, 
noting that the results are not qualitatively changed if they
are varied within reasonable ranges.
For complete technical details of the derivation of 
this effective Hamiltonian and further discussions of higher order terms, as well
as a comparison with other work along similar lines \cite{others}, the
reader is referred to the
supplemental material of Ref. \cite{Bayo}.

\section{Method}

The ground state of the undoped layer is not a simple N{\'e}el-ordered
state. This is due to the spin
fluctuations term $\mathcal{H}_\mathrm{sf}=J_{}/2 \sum_{\langle i,
  j\rangle} (S_{i}^-S_{j}^++S_{i}^+S_{j}^-)$ present in ${\cal H}_{\rm
  AFM}$, which play an important role in lower dimensions. A 2D
solution can only be obtained numerically, for finite size systems
\cite{2DQMC}. The absence of an analytic description of the AFM
background has been an important barrier to understanding what happens
upon doping, because the undoped state itself is so complex. It is
also the reason why most progress has been computational in nature and
mostly restricted to finite clusters. While such results are very
valuable, it can be rather hard to gauge the finite-size effects and,
more importantly, to gain intuition about the meaning of the results.

Because our goal is to verify whether the two kinds of models have equivalent
quasiparticles, which requires us to understand qualitatively what
controls their dynamics, we take a different approach. We use a
quasi-analytic variational method valid for an infinite layer, so that
finite-size effects are irrelevant. By systematically increasing the
variational space we can gauge the accuracy of our guesses and,
moreover, also gain intuition about the importance of various
configurations and the role played by various terms in the
Hamiltonians. Where possible, we compare our results with those
obtained by exact diagonalization (ED) for small clusters, providing
further proof for the validity of our method.

For simplicity, in the following we focus on the one-band model; the
three-band model is treated similarly, as already discussed in
Ref. \cite{Hadi}. Because we do not have an analytic description of
the AFM background wavefunction, we divide the task into
two steps.

\subsection{Quasiparticle in a N\'eel background}

In the first step we completely ignore the spin fluctuations by
setting $\mathcal{H}_\mathrm{sf} \rightarrow 0$, to obtain the
so-called $t$-$t'$-$t''$-$J_z$ model. As a result, the undoped layer
is described by a N\'eel state $|{\rm N}\rangle$ with up/down spins on
the A/B sublattice, without any spin-fluctuations. One may expect this
to be a very bad starting point, given the importance of
spin-fluctuations for a 2D AFM. At the very least, this will allow us
to gauge how important these spin fluctuations really are, insofar as
the quasiparticle dynamics is concerned, when we include them in step
two.

It is also worth remembering that the cuprates are 3D systems with
long-range AFM order stabilized up to rather high temperatures by
inter-layer coupling, in the undoped compounds. The spin fluctuations
must therefore be much less significant in the undoped state than is
the case for a 2D layer, so our starting point may be closer to
reality than a wavefunction containing the full description of the 2D
spin fluctuations.

Magnons (spins wrongly oriented with respect to their sublattice) are
created or removed when the ZRS hops between the two magnetic
sublattices. The creation of an additional magnon costs up to $2J$ in
Ising exchange energy as up to four bonds involving the magnon now become
FM. This naturally suggests the introduction of a variational space in
terms of the maximum number of magnons included in the calculation.

This variational calculation is a direct generalization of that of
Ref. \cite{MonaHolger}, where the quasiparticle of the $t$-$J_z$ model
was studied including configurations with up to 7 {\em adjacent} magnons. That work
showed that keeping configurations with up to three  magnons is
already accurate if $t/J$ is not too large, so here we restrict
ourselves to this smaller variational space. (Note that three is the
minimum number of magnons to allow for Trugman loops \cite{Trugman}, see
discussion below, so a lower cutoff is not acceptable). The configurations
included are the same as  in
Ref. \cite{Trugman}, where this type of approach was first pioneered.

To be more specific, our goal is to calculate the one-hole retarded Green's
function at zero temperature:
\begin{equation}
\label{Gkw}
G({\bf k},\omega) = \langle {\rm N}|d^{\dagger}_{\bf
  k,\uparrow}\hat{G}(\omega)d_{\bf k,\uparrow}|{\rm
  N}\rangle,
\end{equation}
where $\hat{G}(\omega)=\lim_{\eta\rightarrow 0^+}
(\omega-\mathcal{H}+i\eta)^{-1}$ is the resolvent of the one-band Hamiltonian
(\ref{Hamm}) and $d_{\bf k,\uparrow}= \frac{1}{\sqrt{N}}\sum_{i \in
 {\rm Cu}_A}e^{i{\bf k}\cdot{\bf R}_i}d_{i,\uparrow}.$ Here
$N\rightarrow \infty$ is the number of sites in each magnetic sublattice, 
 ${\bf k}$ is restricted to the magnetic Brillouin
zone depicted in Fig. \ref{fig1}(d), and we set $\hbar=1$. The
spectrum is identical if the quasiparticle is located on the
spin-down sublattice: conservation of the total $z$-axis spin
guarantees that there is 
no mixing between these two subspaces with different spin.

Taking the appropriate matrix element of the identity
$\hat{G}(\omega)(\omega + i \eta - \mathcal{H})=1$ leads to the
equation of motion:
\begin{equation}
\label{G1}
[\omega + i \eta -J - \epsilon({\bf k})]G({\bf k},\omega) -t\sum_{\beps
  }F_1({\bf k},\omega, {\beps})=1.
\end{equation}
The four vectors ${\beps= \pm(1,0), \pm(0,1)}$
 point to the nn Cu sites,  $J$ is the Ising exchange energy cost for
 adding the ZRS (four AFM 
bonds are removed) and
\begin{equation}
\label{eps}
\epsilon({\bf k}) = 4t' \cos k_x\cos k_y
+2t''[\cos(2k_x)+\cos(2k_y)]
\end{equation}
 is the kinetic energy of the ZRS moving freely on its own magnetic sublattice. NN
 hopping creates a magnon as the hole moves to the other magnetic
 sublattice; this introduces the one-magnon propagators:
$$F_1({\bf k},\omega, {\beps})=\frac{1}{\sqrt{{N}}}\sum_{i\in
   {\rm Cu}_A}e^{i{\bf k}\cdot{\bf R}_i}\langle {\rm N}|d^{\dagger}_{\bf
   k,\uparrow}\hat{G}(\omega) S^-_id_{i+ {\beps}
   ,\downarrow}|{\rm N}\rangle$$ 
with the hole on the B sublattice and therefore a present magnon on a
nn A site, to conserve the total spin.

Equation (\ref{G1}) is exact (for an Ising background) but it requires
the $F_1({\bf 
  k},\omega, {\beps})\equiv F_1(\beps)$ propagators to solve. Their
equations of motion (EOM) are obtained similarly: 
\begin{multline}
\label{F1}
[\omega + i \eta -{5\over 2}J] F_1( {\beps}) =
 t'\sum_{\beps'\perp \beps}F_1({\beps'})+t''F_1(-{\beps}) \\+t[G({\bf
     k},\omega)+F_2({\beps, \beps})+\sum_{\beps'\perp \beps}F_2({\beps, \beps'})].
\end{multline}
Note that 2$^{\rm nd}$ and 3$^{\rm rd}$ nn hopping keeps the hole on the B
sublattice and thus conserve the number of magnons, linking $F_1$
to other $F_1$ propagators. However, nn hopping links $F_1$ to $G({\bf
  k},\omega)$ if the hole hops back to the
A sublattice by removing the existing magnon, but also to two-magnon
propagators, $F_2$,
if it hops to a different A site than that hosting the first magnon. The equation above imposes the
variational restriction that the two magnons are adjacent, so only  $ 
F_2({\bf k},\omega, {\beps, \beps'})=\sum_{i\in
 {\rm Cu}_A}\frac{e^{i{\bf k}\cdot{\bf R}_i}}{\sqrt{{N}}}\langle {\rm
  N}|d^{\dagger} _{\bf
  k,\uparrow}\hat{G}(\omega) S^-_iS^+_{i+ {\beps}}d_{i+
  {\beps + \beps'} ,\uparrow}|{\rm N}\rangle$ with $\beps + \beps' \ne 0$
are kept. This is a good 
approximation for the low-energy quasiparticle whose magnons are bound
in its cloud, and thus spatially close. (Because
fewer AFM bonds are disrupted, these configurations cost less exchange
energy than those with the magnons apart). Of course, the hole could also
travel far from the first magnon (using 2$^{\rm nd}$ and 3$^{\rm rd}$
nn hopping) before returning to the A sublattice to create a second
magnon far from the first. Such higher energy states---ignored here
but which we consider in the three-band model, see below---contribute  to the
spin-polaron+one magnon continuum which 
appears above the quasiparticle band. The relevance of this
higher-energy feature is
discussed  below.

EOM for the new propagators are generated similarly. We do not write
them here because they are rather cumbersome, but it is clear that the
EOM for $F_2$ link them other $F_2$, as well as 
to some of the $F_1$ and to three-magnon propagators $F_3$. Again we only keep
those propagators consistent with the variational choice of having the 3
magnons on adjacent sites. Since 4-magnon configurations are
excluded, the EOM for $F_3$ link them only to other $F_3$ and to various $F_2$. The resulting
closed system of coupled linear equations is solved numerically to 
find all these propagators, including $G({\bf k},\omega)$.

With $G({\bf k},\omega)$ known, we can find the quasiparticle
dispersion $E({\bf k})$ as the lowest pole of the 
spectral function $A({\bf k}, \omega)=-{1\over \pi} \mbox{ Im} G({\bf
  k},\omega)$. Of course, this is the quasiparticle in a N\'eel
background, {\em i.e.} when the spin fluctuations of the  AFM
background are completely ignored.

\subsection{Quasiparticle in a background with spin fluctuations}

To estimate the effect of the background spin fluctuations (due to
spin flipping of pairs of nn AFM spins, described by
$\mathcal{H}_\mathrm{sf}$) we again invoke a variational
principle. Spin fluctuations occuring far from the ZRS should have no effect
on its dynamics, since they are likely to be ``undone'' before the
hole arrives in their neighborhood (they can be thought of as vacuum
fluctuations).  The spin fluctuations that influence the dynamics of
the hole must be those that occur in its immediate vicinity and either
remove from the quasiparticle cloud pairs of nn magnons generated its
 motion, or add to it pairs of magnons through such AFM
fluctuations. 

For consistency, we keep the same variational configurations here like
we did at the previous step. Then, Eq. (\ref{G1}) acquires an
additional term on the l.h.s. equal to: $-{J\over 2}
\sum_{{\beps+\beps'}\ne 0} e^{i{\bf k}\cdot ({\beps+\beps'})}F_2({\bf
  k}, \omega, {\beps, \beps'})$, describing processes where a pair of
magnons is created through spin-fluctuations near the
hole. Similarly, the EOMs for $F_1$/$F_2$/$F_3$ acquire terms
proportional to $F_3$/$G$/$F_1$ respectively, because spin
fluctuations add/remove a pair of magnons to/from their clouds. This
modified system of linear equations has a different solution for
$G({\bf k}, \omega)$, which accounts for the effects of the spin
fluctuations that occur close to the hole. Comparison with the
previous results will allow us to gauge how important these
``local'' spin fluctuations are to the quasiparticle's dynamics.

Accuracy can be systematically improved by increasing the variational
space, and implementation of such generalizations is
straightforward. As shown next, the results from the variational
calculation with configurations of up to three magnons located on
adjacent sites compares well against available ED results and allows
us to understand what determines the quasiparticle's dispersion, so we
do not need to consider a bigger variational space.

The three-band model is treated similarly, with the variational space
again restricted to the same configurations with up to three adjacent
magnons. Results for a quasiparticle in the N\'eel background (no spin
fluctuations) were published in Ref. \cite{Hadi}, where the reader can
find details about the corresponding EOMs (see also
Ref. \cite{Hadithesis}). Here we will focus primarily on the effect of
local spin fluctuations. These are introduced as explained above, by
adding to the EOMs terms consistent with the variational space and
whose magnon count varies by two.

\section{Results}

\subsection{One-band model}

We first present results for the one-band model. It is important to
note upfront that it has long been known that both spin fluctuations
and the longer range hopping must be included in order to obtain the
correct dispersion for its quasiparticle \cite{Leung95,Leung97}. (By
``correct dispersion'' we mean one in agreement with the experimental data
\cite{Wells,Damascelli}). Our results confirm these facts, as shown next.

The novelty is, therefore, not in finding these results, but in using
them to prove the validity of our variational method and, more
importantly, to untangle the specific role that spin fluctuations and
long-range hopping play in arriving at this dispersion. To the best of
our knowledge, this had not been known prior to this work.

\begin{figure}[t]
\center \includegraphics[width=.99\columnwidth]{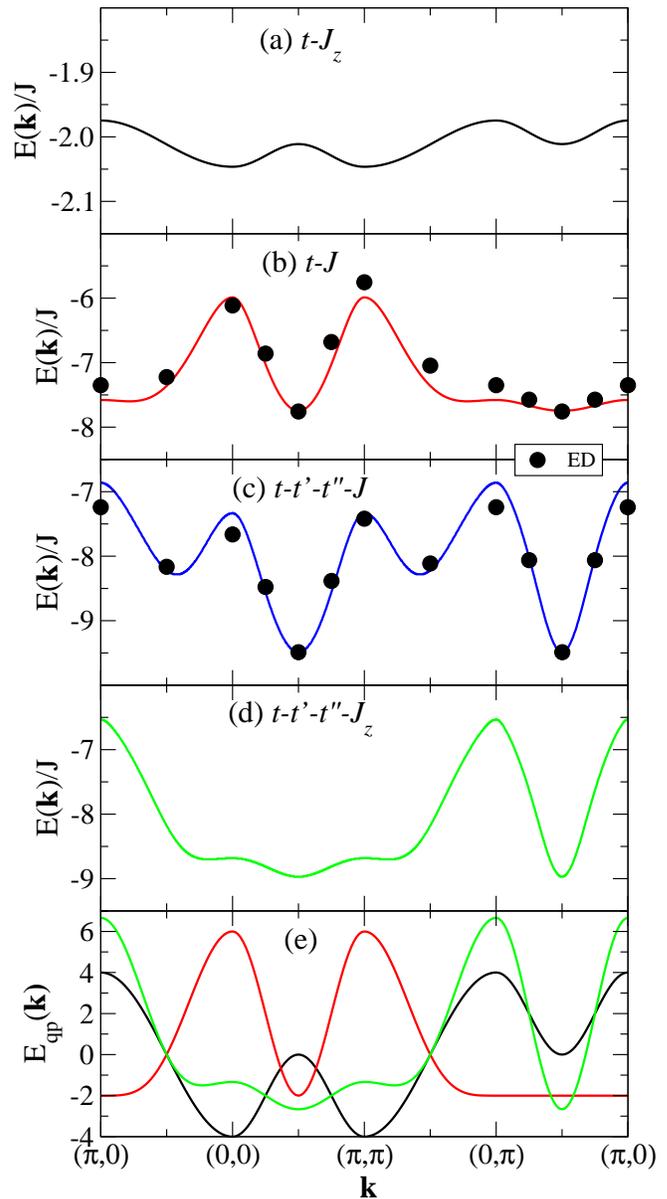}
\caption{(color online) Quasiparticle energy $E({\bf k})$ along several cuts
  in the Brillouin zone for various one-band models. In all cases
  $J/t=0.3$ while $t'=t''=0$ in 
  (a),(b), and $t'/t=-0.3, t''/t=0.2$ in (c),(d). Lines
  show the results of the variational calculation with the spin
  fluctuations frozen in (a) and (d), or allowed only near the hole
  in (b) and (c). Symbols in (b) and (c) are the corresponding ED results for a 
  32-site cluster \cite{Leung95,Leung97}. (e) Dispersion
  $E_{qp}({\bf k)}$ of Eq. (\ref{disp}) for $E_0=0$ and $t_2=-1, t_3=0$ (black);
  $t_2=1, t_3=0.5$ (red); $t_2=-1; t_3=2/3$ (green).  }
\label{fig3}
\end{figure}

The quasiparticle dispersion  $E({\bf k})$ is shown in
Fig. \ref{fig3}  for various models: in panel (a) we set $t'=t''=0$ and freeze
spin-fluctuations ($t$-$J_z$ model). In panel (b), spin fluctuations
close to the hole are turned on as discussed; for simplicity we call
this the $t$-$J$ model, although the true $t$-$J$ model
includes all spin fluctuations. In panel (c), we further add the
longer range hopping; for simplicity, we call this the
$t$-$t'$-$t''$-$J$ model although, again, spin fluctuations are
allowed only near the hole. Finally, in panel (d) we keep the longer
range hopping but freeze the spin fluctuations; this is the
$t$-$t'$-$t''$-$J_z$ model. Panel (e) shows model dispersions
explained below.

The quality of our variational approximation is illustrated in panels
(b) and (c). Its results (thick lines) are in fair agreement with
those of exact diagonalization (ED) for a 32-site cluster, which
includes all spin fluctuations \cite{Leung95,Leung97}. Results in
panel (c) agree well with those measured experimentally
\cite{Leung97,Damascelli}. Our bandwidths are somewhat different; some
of this may be due to finite-size effects, as the ED bandwidth varies
with cluster size \cite{Bayo}. This also suggests that more
configurations need to be included before full convergence is
reached by our variational method (these would increase the bandwidth
in panel (b) and decrease it in panel (c), see below). This is
supported by Ref. \cite{MonaHolger}, where full convergence for the
$t$-$J_z$ model was reached when configurations with up to 5 magnons
were included. Nevertheless, the agreement is sufficiently good to
conclude that the essential aspects of the quasiparticle physics are
captured by the three-magnon variational calculation, and to confirm
that it suffices to include spin fluctuations only near the hole.

\begin{figure*}[t]
\center \includegraphics[width=1.99\columnwidth]{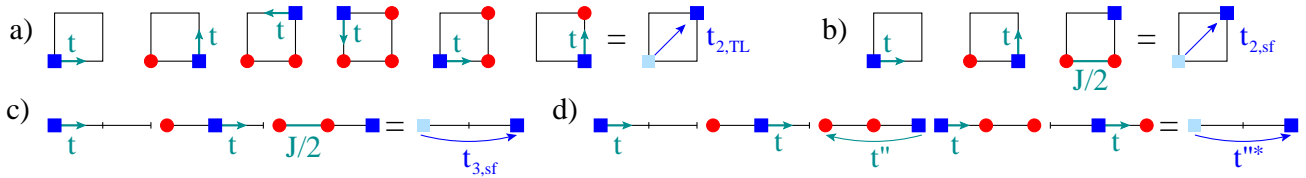}
\caption{ (color online) (a) Shortest Trugman loop that generates a
  $t_{2, \rm TL}$ contribution. A much smaller $t_{3, \rm TL}$ is also
  generated, see Ref. \cite{MonaHolger}; Generation of (b) $t_{2, \rm sf}$ and
  (c) $t_{3, \rm sf}$ terms due to spin-fluctuations; (d) Process that
  renormalizes $t''$. The square shows the location of the hole, while
  the circles show magnons (wrongly oriented spins). The remaining
  spins are in their N\'eel order orientation and are not shown
  explicitly. The short thick arrows indicate the next step in the
  process, while the thin arrows in the final sketch show the
  effective quasiparticle hopping generated by those processes. }
\label{fig4}
\end{figure*}

These results clearly demonstrate that both spin fluctuations and
longer-range hopping are needed to achieve the correct
quasiparticle dispersion shown in panel (c), with deep, nearly isotropic
minima at $({\pi\over 2}, {\pi\over 2})$. Absence of longer-range
hopping leads to a rather flat dispersion along $(0,\pi)-(\pi,0)$, see
panel (b); this fact has long been known \cite{Leung95}. In panel (d),
we show that if the longer range hopping is included but spin
fluctuations are absent, the dispersion is rather flat along the
$(0,0)-(\pi,\pi)$ direction. We are not aware of previous studies of this
case.

Although $E({\bf k})$ looks very different in the four cases, it turns
out that all can be understood in a simple, unified picture. The key
insight is that the quasiparticle lives on one magnetic sublattice,
because of
spin conservation. As a result, its generic dispersion must be of
the form:
\begin{equation}
\label{disp}
E_{qp}({\bf k}) = E_0+ 4t_2 \cos k_x \cos k_y + 2 t_3 ( \cos 2k_x + \cos 2k_y)
\end{equation}
{\em i.e.} like the bare hole dispersion $\epsilon({\bf k})$ of
Eq. (\ref{eps}), but with renormalized 2$^{\rm nd}$ and 3$^{\rm rd}$
nn hoppings $t'\rightarrow t_2; t'' \rightarrow t_3$. There cannot be
any effective nn hopping of the quasiparticle because this would move
it to the other sublattice; this cannot happen without changing the
magnetic background, so $t_1=0$.  Longer
range hoppings that keep the quasiparticle on the same sublattice may
also be generated dynamically, but their magnitude is expected to be
small compared to $t_2,t_3$, hence Eq. (\ref{disp}). Thus,  understanding the
shape of the quasiparticle dispersion requires understanding
the values  of $t_2$ and $ t_3$.

We begin the analysis with the $t$-$J_z$ model. Its quasiparticle is
extremely heavy, as shown in panel (a). Note that the vertical scale
is an order of magnitude smaller than for the other panels.  The
reason is that every time the hole hops, it moves to the other magnetic
sublattice and it must either create or remove a magnon, to conserve
the total spin. As the hole moves away from its original location, it
leaves behind a string of magnons whose energy increases roughly
linearly with its length. This could be expected to result in
confinement (infinite effective mass), but in fact the quasiparticle
acquires a finite dispersion by executing Trugman loops (TL)
\cite{Trugman}, the shortest of which is sketched in
Fig. \ref{fig4}(a). By going nearly twice along a closed loop,
creating a string of magnons during the first round and removing them
during the second round, the hole ends up at a new location on the same
magnetic sublattice.  Only 2$^{\rm nd}$ and 3$^{\rm rd}$ nn hopping
terms can be generated through TL irrespective of their length, and
$|t_{3,\rm TL}|\ll |t_{2,\rm TL}|\ll J$ if $t/J \sim 3$
\cite{MonaHolger}.  Indeed, setting $t_2 <0, t_3\rightarrow 0$ in
$E_{qp}({\bf k})$ of Eq. (\ref{disp}) leads to the black curve in
Fig. \ref{fig3}(e) \cite{c1}, which has the same shape as that of
panel (a) (the bandwidth is proportional to $|t_{2,\rm TL}|$). This
dispersion is wrong not just quantitatively but also qualitatively,
with $({\pi\over 2}, {\pi\over 2})$ as a saddle point instead of the
ground state. Clearly, ignoring both longer range hopping and spin
fluctuations changes completely the dynamics of the quasiparticle.

When the spin fluctuations are turned on in the $t$-$J$ model, they
act on a time scale $\tau_{\rm sf} \sim 1/J$ much faster than the slow
dynamics due to TL, $\tau_{\rm TL} \sim 1/|t_{2,\rm TL}|$. The main
contributions to $t_2$ and $t_3$ now come from processes like those
sketched in Fig. \ref{fig4}(b) and (c), where spin fluctuations remove
pairs of magnons created by nn hopping of the hole, leading to
$t_{2,{\rm sf}}\gg |t_{2, \rm TL}|, t_{3,\rm sf}\gg |t_{3, \rm
  TL}|$. Moreover, we expect $t_{2,\rm sf}= 2 t_{3,\rm sf}$
because these effective hoppings are generated by similar processes
but there are twice as many leading to 2$^{\rm nd}$ compared to 3$^{\rm
  rd}$ nn hopping, as the hole can move on either side of a plaquette.

Indeed, the $t$-$J$ dispersion of panel (b) has a shape similar to
that of 
$E_{qp}({\bf k})$ with $t_2=2t_3$, shown as a red curve in
Fig. \ref{fig3}(e) \cite{c2}. Because $E_{qp}(k, \pi-k) = E_0 - 2t_2 +
(4t_3-2t_2)\cos 2k$, it has a perfectly flat dispersion along
$(0,\pi)-(\pi,0)$ for $t_2=2t_3$. The dispersion in panel (b) is not
perfectly flat along this cut, so in reality $t_{2} \approx 2
t_{3}$. The small correction from the factor of 2 is likely due to
higher order processes, as well as contributions from TL (which remain
active). Ignoring it, we find the corresponding bandwidth
$E_{qp}(0,0)-E_{qp}({\pi\over 2}, {\pi\over 2}) = 4t_2+8t_3=8t_2$,
suggesting that the effective hoppings generated with spin
fluctuations are of the order $t_{2,\rm sf}\approx 2 t_{3,\rm
  sf}\approx J/4$.

Next, we consider what happens if instead of (local) spin
fluctuations, we turn on longer-range hopping.  Unlike in the
$t$-$J_z$ model, the quasiparticle of the $t$-$t'$-$t''$-$J_z$ model
should be light because the longer range hoppings $t', t''$ allow the
hole to move freely on its magnetic sublattice. It can therefore
efficiently remove magnons created through its nn hopping, without
having to complete the time-consuming Trugman loops. The presence of
the magnon cloud renormalizes these bare hoppings to smaller values,
as is typical for polaron physics. Figure \ref{fig4}(d) shows one such
process that renormalizes $t''\rightarrow t''^*$. Similar processes
(not shown) renormalize $t'\rightarrow t'^*$ so both hopping integrals
should be renormalized by comparable factors. As a result, we expect a
dispersion like $E_{qp}({\bf k})$ but now with $t_2/t_3 = t'^*/t''^*
\approx t'/t''$, if we ignore the small TL contributions. This indeed
agrees with the result in panel (d), as shown by its comparison with
the green curve in Fig. \ref{fig3}(e) where $E_{qp}({\bf k})$ is
plotted for $t_2/t_3 = t'/t''= -1.5$. For $t_2=-2t_3$, $E_{qp}({\bf
  k})$ would be perfectly flat along $(0,0)-(\pi,\pi)$. Thus, the
change in the relative sign explains why now the dispersion is nearly
flat along $(0,0)-(\pi,\pi)$ and maximal along $(0,\pi)-(\pi,0)$, in
contrast to the previous case. However, while $t_{2,\rm sf}/t_{3,\rm
  sf}\approx$ 2 is always expected for the $t$-$J$ model so its
dispersion must have a shape like in Fig. \ref{fig3}(b), in the
$t$-$t'$-$'t''$-$J_z$ model the ratio $t'^*/t''^*$ mirrors the
ratio $t'/t''$. If this had a very different value than $\approx -2$, the
quasiparticle dispersion would change accordingly.

These results allow us to understand the dispersion of the
$t$-$t'$-$t''$-$J$ quasiparticle. This must have contributions from
both the spin fluctuations and the renormalized longer range hoppings,
plus much smaller TL terms, because the processes giving rise to them
are now all active. Indeed, the curve in panel (c) of Fig. \ref{fig3}
is roughly equal to the sum of those in panels (b) and (d). The
isotropic minimum at $\left({\pi\over2}, {\pi\over2}\right)$ is thus
an accident, since the dispersion along $(0,0)-(\pi,\pi)$ is
controlled by spin fluctuations, and that along $(0,\pi)-(\pi,0)$ is
due to the renormalized longer range hoppings. More precisely, because
$t_{2,\rm sf} \approx 2 t_{3,\rm sf}$, the contributions coming from
spin fluctuation interfere destructively for momenta along
$(0,\pi)-(\pi,0)$ so dispersion here is controlled by the renormalized
$t'^*\approx -1.5 t''^*$, and viceversa. If $t_{2,\rm sf} \approx
|t'^*|$ (which happens to hold because $J \sim |t'|$), the sum gives
nearly isotropic dispersion near $\left({\pi\over2},
{\pi\over2}\right)$. If we change parameters significantly, the
dispersion becomes anisotropic (not shown).

\begin{figure}[t]
\includegraphics[width=\columnwidth]{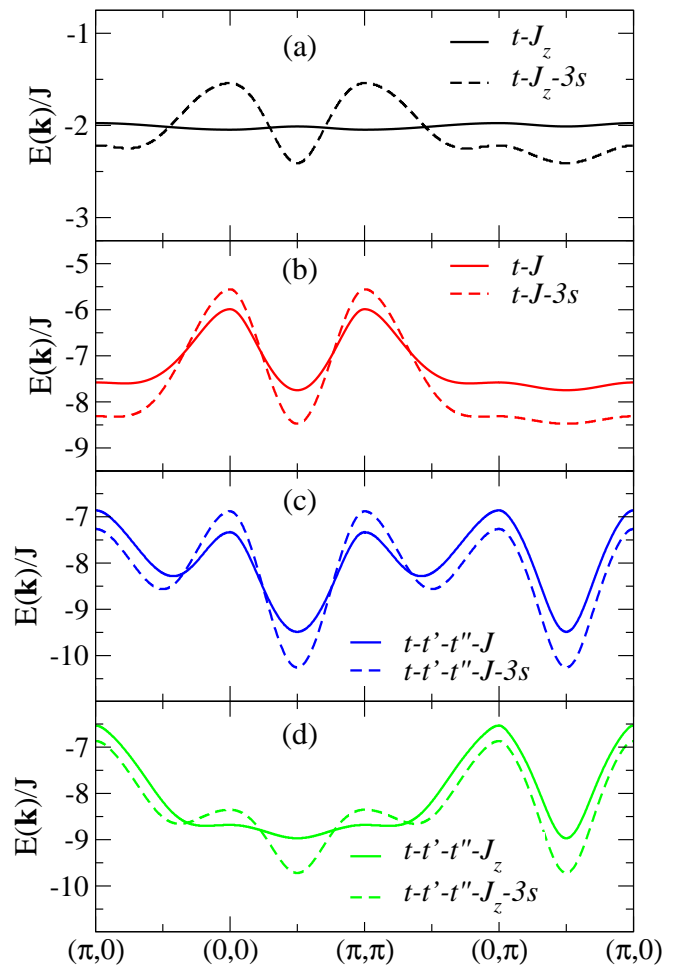}
\caption{ (color online) Quasiparticle dispersion  when the three-site
  term of Eq. (\ref{3site}) is included in 
  the one-band Hamiltonians (dashed lines). For comparison, the dispersions
  without this term are also show (full lines from Fig. \ref{fig3}) }
\label{fig5}
\end{figure}

Before moving on to contrast this behavior with that of the
quasiparticle of the three-band model, we briefly discuss the effect
of the three-site term of Eq. (\ref{3site}). The variational results
for the four models are shown in Fig. \ref{fig5}. Where direct
comparisons can be made, they are again in good quantitative agreement
with other work where this term has been included, such as in
Ref. \cite{Bala}.  Its inclusion has a qualitative effect only for the
$t$-$J_z$ model, where the shape of the dispersion is changed in its
presence. This is not very surprising because, as discussed, the
Trugman loops which control behavior in that case are very slow
processes, and their effect can easily be undone by terms that allow
the hole to move more effectively. The three-site term is such a term
and its presence increases the bandwidth not just for the $t$-$J_z$
model, but for all cases.  For the other three models, however, the
inclusion of this term changes the dispersion only quantitatively: the
bandwidth is increased but the overall shape is not affected much. The
biggest change is along $(0,0)-(\pi,\pi)$, as expected because the
three-site term generates effective 2nd and 3rd nn hoppings with the
{\em same sign} and a 2/1 ratio, {\em i.e.} similar to $t_{2,sf}$ and
$t_{3,df}$. As a result, its presence mimics (and boosts) the effect of the local
spin fluctuations.

It is interesting to note that if we allow this term
to be large enough, we could obtain a dispersion with the correct
shape {\em even in the absence of spin fluctuations}. However, the
scale of this term is set by $J$, it is not a free parameter. As a
result, we conclude that with its proper $J$ energy 
scale, this terms
does not change qualitatively the behavior of the quasiparticle of the
one-band model (apart from the $t$-$J_z$ case), although its inclusion
may, in principle, allow for better fits of the experimental data.

\subsection{Three-band model}

\begin{figure}[t]
\center \includegraphics[width=\columnwidth]{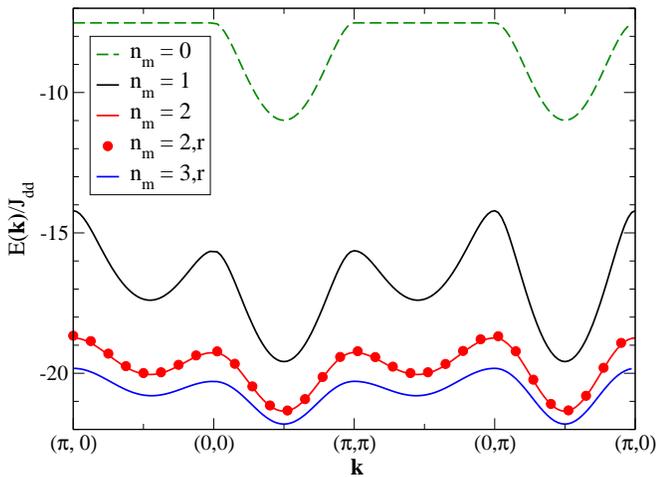}
\caption{ (color online) $E({\bf k})$ along several cuts in the
  Brillouin zone for the three-band model. The results are for the
  variational calculation with the spin-fluctuations turned off and
  configurations with up to $n_m$ magnons allowed. The ``restricted''
  calculations labelled $n_{2,r}, n_{3,r}$ imposed the additional
  constraint that the magnons are on adjacent sites. While the
  bandwidth is strongly renormalized with increasing $n_m$, the nearly
  isotropic dispersion around the ground-state at $({\pi\over 2},
  {\pi\over2})$ is a consistent feature. See text for more details.}
\label{fig6}
\end{figure}

Results for the simplified three-band model with the spin-fluctuations
frozen off were discussed in Ref. \cite{Hadi}. To keep this work
self-contained, we show in Fig. \ref{fig6} the most relevant data for
the issue of interest, namely the
quasiparticle dispersion $E({\bf k})$ obtained  in a variational calculation with
the maximum number of  magnons $n_m = 0 - 3$. These results already
suffice to illustrate the qualitative difference between the
quasiparticle dynamics in the one-band
and the three-band models.

The $n_m=0$ curve  plots the dispersion
if no magnons are allowed, {\em i.e.} not only the spin
fluctuations of the AFM background but also spin-flip
processes due to $J_{pd}$ and $T_{\rm swap}$ are turned off. It is
important to emphasize that the resulting dispersion does 
contain a very important contribution from the terms in $T_{\rm swap}$ describing
hopping of the hole past Cu spins with the same spin projection, so
that the spin-swap leaves the spins unchanged.

In fact, it is the interference between these terms with those of
$T_{pp}$ that leads to this interesting bare dispersion, which already
has deep, nearly-isotropic minima near $({\pi\over 2},
{\pi\over2})$. If we set $t_{sw}=0$, the bare dispersion due to only
$T_{pp}$ has the ground-state at $(\pi, \pi)$, whereas if $T_{pp}=0$,
the dispersion due to the allowed terms in $T_{\rm swap}$ is perfectly
flat because the hole is then trapped near a like Cu spin. However, as long
as $t_{pp} \sim t_{sw}$, the isotropic minimum emerges at $({\pi\over
  2}, {\pi\over2})$. In this context, it is useful to note that in
many numerical studies of the Emery model, $t_{pp}$ was set to zero
simply for convenience.  Our results suggest that this choice changes
 the quasiparticle dynamics qualitatively, and is therefore unjustified.

The bare hole dispersion in the three-band
model thus already mimics this key aspect of the correct quasiparticle
dispersion, unlike in the one-band model. Allowing the quasiparticle
cloud to emerge by allowing the 
hole to create and absorb magnons in its vicinity, through spin-flip processes
controlled by $J_{pd}$ and $T_{\rm swap}$, further renormalizes the
bandwidth (a typical polaronic effect) without affecting the existence
of the isotropic dispersion near $({\pi\over 2}, {\pi\over2})$.  This
magnon cloud is very important, however, to stabilize the low-energy
quasiparticle, as demonstrated by the significant lowering of the
total energy. In particular, at least one magnon must be present in
order for a ZRS-like object to be able to form, and indeed the
$n_m=1$ curve is pushed down by $\sim 10 J_{dd}$ compared to the bare
dispersion. We further analyze the relevance of the ZRS solution
below.

The small difference between the $n_m=2$ and $n_m=2,r$ results proves
that magnons indeed sit on adjacent sites in the cloud. (The latter
solution imposes this constraint explicitly, whereas the former allows
the magnons to be at any distance from each other. In both cases, the
hole can be arbitrarily far from the magnons, although, as expected,
configurations where the hole is close to the last emitted magnon have
the highest weight in the quasiparticle eigenstates.) At higher
energies, however, these two solutions are qualitatively
different. The former contains the expected quasiparticle+one-magnon
continuum starting at $E_{1, gs}+2J_{dd}$, where $E_{1,gs}$ is the
ground-state energy of the quasiparticle with $n_m=1$, and $2J_{dd}$
is the energy cost to create a magnon far from it. Their sum is the
energy above which higher-energy (excited) states must appear in the
spectrum, describing the quasiparticle plus one magnon not bound to
its cloud. The presence of this continuum guarantees that in the fully
converged limit, the quasiparticle bandwidth cannot be wider than
$2J_{dd}$, since the quasiparticle band is always ``flattened out''
below this continuum (another typical polaronic behavior). For both
$n_{2,r}$ and $n_{3,r}$ calculations, the quasiparticle is already
heavy enough that its dispersion fits below the corresponding
continuum. This is why enlarging the variational space with
configurations needed to describe this feature, with at least one
magnon located far from the cloud, does not affect the quasiparticle
dispersion much (see Ref. \cite{Hadi} for more discussion).

The bandwidth of the $n_m=3,r$ dispersion is in decent agreement with
numerical results for this model, as discussed next, suggesting that
this variational calculation is close to fully converged. The fact
that the cloud is rather small should not be a surprise. The
variational approach explicitly imposes the constraint that there is
at most one magnon at a site.  As magnons sit on adjacent sites when
bound in the quasiparticle cloud, they prefer to occupy a compact area
to minimize their exchange energy cost, thus creating a domain in the
other N\'eel state (down-up instead of up-down). The hole prefers to
sit on the edge of this domain, because being inside it is equally
disadvantageous to being outside, {\em i.e.} far from
magnons. However, since on the boundary the hole can interact with
only one magnon at one time, a large and costly domain is unlikely.

We can now contrast the dynamics of the quasiparticle in the
three-band model {\em if the spin fluctuations are frozen out} with
the corresponding one-band model, namely the $t$-$t'$-$t''$-$J_z$ case. Both
have a quasiparticle with a small, few-magnon cloud,
and a bandwidth  $\approx 2 J = 2J_{dd}$. The key difference is
that the three-band model already shows a dispersion with the correct
shape, whereas for the one-band model the dispersion is much too flat
along $(0,0)-(\pi,\pi)$. This difference is traced back to the fact that
in the one-band model, the bare hole dispersion also suffers from this
same problem if 
$t'/t''\sim -1.5$, unlike
that of the three-band model. As a result, spin
fluctuations are necessary to find the correct
dispersion in the one-band model, as already shown, but their role
in the three-band model should be rather limited.

\begin{figure}[t]
\center \includegraphics[width=\columnwidth]{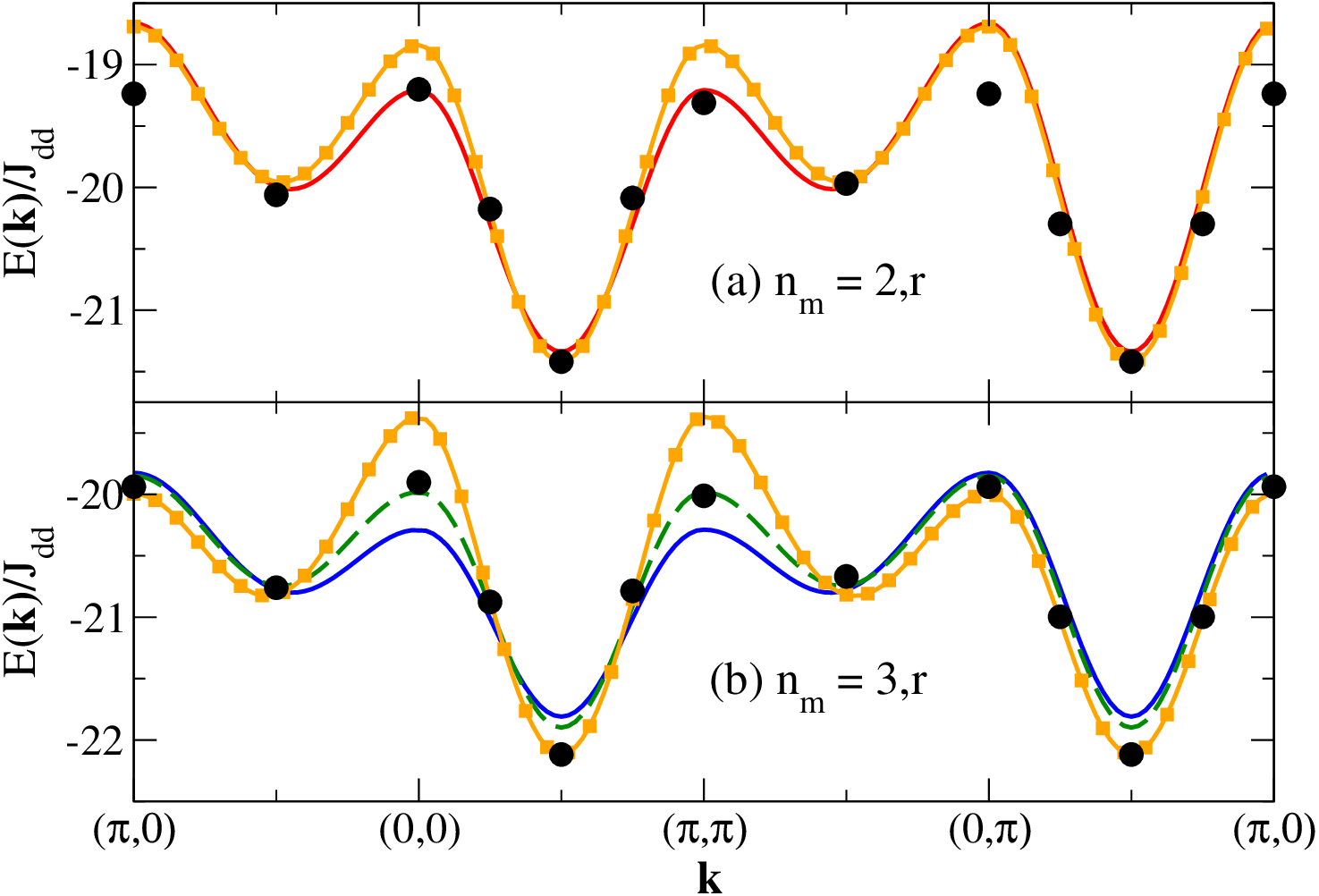}
\caption{ (color online) $E({\bf k})$ along several cuts in the
  Brillouin zone for the three-band model in the restricted
  variational approximations with (a) $n_m=2$ and (b) $n_m=3$. Circles show 
  ED results for $S_T={1\over 2}$ from Ref. \cite{Bayo} for a  32 Cu +
  64 O cluster, shifted to have the same ground-state energy. Full
  lines show the results of Fig. \ref{fig6}, without spin
  fluctuations. Orange lines with square symbols are the results if 
  spin fluctuations occur near the hole. The dashed green line in
  panel (b)  is the dispersion when spin fluctuations are allowed
  to locally create/remove a pair of magnons only if no other magnons are 
  present/remain in the system. See text for more details.}
\label{fig7}
\end{figure}

To confirm this conjecture, we consider the effect of local spin
fluctuations on the dispersion of the three-band model
quasiparticle. In Fig. \ref{fig7}, results for the restricted
variational approach with $n_m=2,r$ and $n_m=3,r$ ({\em i.e.} up to
two or up to three magnons on adjacent sites) are compared to the ED
results of Ref. \cite{Bayo}, shown by the black full circles. Full
lines (red and blue, respectively) show the results of
Fig. \ref{fig6}, without spin fluctuations. Orange lines with squares
show the dispersion with spin fluctuations turned on near the
hole. The dashed green line in panel (b) shows an intermediate result
when spin fluctuations are allowed to create a pair of magnons only if
there is no magnon in the system, and to remove a pair if only two
magnons are present (the orange line also includes contributions from
processes where spin fluctuations add a pair of magnons when a magnon
is already present, and its reversed process).

The effect of spin fluctuations is similar to that
found in the one-band models, as expected because the AFM
background is modeled identically. They again have very
little effect on the $(\pi,0)-(0,\pi)$ dispersion; beside a small
shift to lower energies, this bandwidth is only slightly increased,
bringing it into better agreement with the ED values for $n_m=3$. Like
for one-band models, spin fluctuations lead to a more significant
increase of the $(0,0)-(\pi,\pi)$ dispersion. For $n_m=3$, it changes
from being too narrow without spin fluctuations, to too wide in
their presence. (The $n_m=2$ overestimate of the bandwidth is
expected, see discussion in Ref. \cite{Hadi}).

The increased energy near $(0,0) = (\pi,\pi)$ may seem problematic but
one must remember that in reality, $E({\bf k})$ is flattened below a
continuum that appears at $2 J_{dd}$ above the ground state. The
continuum is absent in this restricted calculation because
configurations with a magnon far from the cloud, which give rise to
it, are not included. This explains why the overestimated bandwidth is
possible. In the presence of the continuum, states that overlap with
it hybridize with it and a discrete state (the quasiparticle) is
pushed below its edge.  This will lower the value at $(0,0)$ and lead
to good quantitative agreement everywhere with the ED results. (The
keen reader may note that the ED bandwidth is also slightly wider than
$2J_{dd}$, but one must remember both the finite size effects of
cluster ED, and the existence of a $S_T={3\over 2}$ polaron in its
spectrum \cite{Bayo}. Since the total spin ${\hat {\bf S}}_T^2$ is not
a good quantum number in our variational approximation, its
quasiparticle probably overlaps somewhat with both the $S_T={1\over
  2}$ and $S_T={3\over 2}$ spin-polarons, so it is not clear which ED
states to compare against).

Our results show that spin fluctuations have a similar effect in
both models. However, while they are essential for restoring the
proper shape of the dispersion in one-band models, they are much less
 relevant for the three-band model. This is a direct
consequence of the different shape of the bare bands, as discussed,
but also to having $J\sim |t'|$ while $J_{dd} \sim t_{pp}/4$. In the
three-band model, the quasiparticle creates and absorbs magnons while
moving freely on the O sublattice, on a timescale that is faster than
that over which spin fluctuations act, and so their effect is
limited. In contrast, in the one-band model, the timescale for free
propagation of the hole on the same magnetic sublattice (controlled by
$t', t''$)  is comparable
with the spin fluctuations' timescale, and therefore the effect of
spin fluctuations is much more significant. They are especially
important along $(0,0)-(\pi,\pi)$, where the bare dispersion
of one-band models is
nearly flat.

\section{Discussion and summary}

In this work, we used a variational method to study and compare the
quasiparticle of  $t$-$J$ and $t$-$t'$-$t''$-$J$  one-band
models, to that of a (simplified) three-band model that is the
intermediary step between the full three-band Emery model and the
one-band models.

Our variational method generates the BBGKY hierarchy of equations of
motions for a propagator of interest (here, the retarded one-hole
propagator), but simplified by setting to zero the generalized
propagators related to projections on states that are not within the variational
space. Its physical motivation is very simple: if the variational
space is properly chosen, {\em i.e.} if it contains the configurations
with the highest weight contributions to the quasiparticle
eigenstates, then the ignored propagators are indeed small because
their residue at the $\omega =E({\bf k})$ pole is proportional to
their weight (Lehmann representation). Setting them to zero should
thus be an accurate approximation. Numerically, the motivation is
also clear: because the resulting simplified hierarchy of coupled
equations can be solved efficiently, we can quite easily study a quasiparticle  (or
a few \cite{MonaPRL}) on an infinite plane, thus avoiding
finite-size effects and getting full information about ${\bf k}$
dependence, not just at a few values. Moreover, by enlarging the
variational space and by turning off various terms in the Hamiltonian,
both of which lead to changes in the EOM and thus the resulting
propagators, one can infer whether the calculation is close to
convergence and isolate and understand the effect of various terms,
respectively. The ability to efficiently make such comparisons is
essential because it allows us to gain intuition about the resulting
physics.

Our results show that even though  for reasonable values of the
parameters, the quasiparticle dispersion $E({\bf
  k})$ has similar shapes in both models, the underlying quasiparticle
dynamics is very 
different. In the three-band model, the bare dispersion of the hole on
the O sublattice, due to $t_{pp}$ and spin-swap hopping $t_{\rm sw}$
past Cu with parallel spins, already has a deep isotropic minimum near
$\left({\pi\over2}, {\pi\over2}\right)$, unlike the bare 
$\varepsilon({\bf k})$ of the one-band models. When renormalized due to
the magnon cloud, it produces a quasiparticle dispersion with the
correct shape in the whole Brillouin zone, even in the absence of spin
fluctuations. In contrast, for the one-band models the inclusion of
spin fluctuations is necessary for the correct dispersion to
emerge. This shows that the quasiparticle dynamics is controlled by
different physics in the two models, and this is likely to play a role
at finite concentrations as well. 

Our results thus raise strong doubts on whether the one-band
$t$-$t'$-$t''$-$J$  model
truly describes the same physics like the three-band model. We can
think of three possible explanations to explain these differences:

(i) The $t$-$t'$-$t''$-$J$ model is the correct one-band model, but its true
parameters have values quite different from the ones used
here. Indeed, if the bare $\varepsilon({\bf k})$ dispersion had isotropic minima at
$({\pi\over2}, {\pi\over 2})$, and if its renormalized bandwidth would
be of the order of $2J$, then spin fluctuations could not change it
much, similar to what is observed in the three-band model.

This explanation can be ruled out. An isotropic bare dispersion
$\varepsilon(k,k) \approx \varepsilon(k,\pi-k)$ requires that
$|t''/t'| \gg 1$, which is physically unreasonable.

(ii) The $t$-$t'$-$t''$-$J$ model has a different quasiparticle
because its underlying assumption, {\em i.e.} the existence of the
low-energy ZRS, is wrong. This would mean that not only
this specific one-band model but any other one obtained through such a
projection would be invalid.

\begin{figure}[t]
\center \includegraphics[width=\columnwidth]{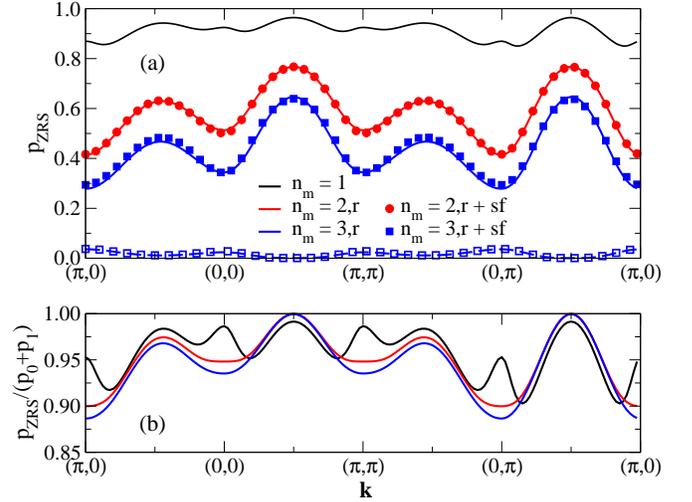}
\caption{ (color online) (a) Overlap $p_{\rm ZRS}$ between the ZRS Bloch state of Eq. (\ref{zrs}) and the quasiparticle
eigenstate, as obtained in the restricted variational
calculations with $n_m=1,2,3$, without (full lines) and with
(full symbols) local spin fluctuations included. The empty squares and
the dashed line show the weight of the ZR triplet. (b) Overlap $p_{\rm ZRS}$ normalized with respect to the probability to have no magnon ($p_0$) or to have one magnon near the hole ($p_1$), in the quasiparticle eigenstate. See text for more
details. }
\label{fig8}
\end{figure}

We can test this hypothesis by calculating the
overlap between the three-band quasiparticle and a ZRS Bloch state.
The latter is defined in the only possible way that is consistent with
 N\'eel order: 
\begin{equation}
\label{zrs}
|{\rm ZRS},{\bf k}\rangle = {1\over \sqrt{N}} \sum_{i\in {\rm
    Cu}_{\downarrow} }^{}e^{i {\bf k} \cdot {\bf R}_i} {
  p^\dagger_{x^2-y^2, i, \uparrow} - p^\dagger_{x^2-y^2, i,
    \downarrow}S_i^{+}\over \sqrt{2}} |{\rm N} \rangle
\end{equation}
where
$$ p^\dagger_{x^2-y^2, i, \sigma}={1\over 2}
\left[p^\dagger_{i+{{\bf x}\over 2},\sigma } + p^\dagger_{i+{{\bf
      y}\over 2},\sigma} - p^\dagger_{i-{{\bf x}\over 2}, \sigma}-
  p^\dagger_{i-{{\bf y}\over 2}, 
 \sigma   } \right]
$$ is the linear combination of the $p$ orbitals neighbor to the Cu
located at $i$, that has the overall $x^2-y^2$ symmetry (our choice
for the signs of the lobs is shown in Fig. \ref{fig1}).
Note that with this definition  $|\langle {\rm ZRS},{\bf k}|
{\rm ZRS},{\bf k}\rangle|^2=1$ for any 
${\bf k}$, so there are no normalization
problems \cite{Zhang}.

We define $p_{\rm ZRS} = |\langle qp, {\bf k}| {\rm ZRS},{\bf
  k}\rangle|^2$ as the overlap between the quasiparticle eigenstate of
momentum ${\bf k}$ and this ZRS Bloch state. Its value can be
calculated from the appropriate 
residues of the zero- and one-magnon propagators at $\omega = E({\bf
  k})$, and is shown in Fig. \ref{fig8}(a). We do not plot the $n_m=0$
results because a singlet cannot form if the Cu spins cannot
flip. (For $n_m=0$, there is overlap with the spin-up hole component
of $|{\rm ZRS},{\bf k}\rangle$, and we find that $p_{\rm ZRS}$ varies
from 0 at $(0,0)$ and $(0,\pi)$ to 0.5 at $({\pi\over2},{\pi\over2})$, but
the same answer is found for a triplet. Interestingly, this proves
that the bare hole dispersion already has eigenstates with the
$x^2-y^2$ symmetry near $({\pi\over2}, {\pi\over2})$).

For $n_m=1$ we find $p_{\rm ZRS} \sim 0.9$ in the entire Brillouin
zone. Clearly, in this very small variational space, locking into a
ZRS is the best way for the doped hole to lower its energy.  However,
the value of $p_{\rm ZRS}$ decreases fairly significantly for
$n_m=2,r$ and $n_m=3,r$. First, note that turning the spin
fluctuations on or off has almost no effect on $p_{\rm ZRS}$. This is
consistent with our conclusion that local spin fluctuations do not
influence the nature of the quasiparticle in the three-band model:
clearly, its wavefunction is not changed in their presence.

The decrease of $p_{\rm ZRS}$ with increasing $n_m$ could be due
either to increased contributions to the eigenstate from many-magnon
configurations (which have no overlap with $|{\rm ZRS},{\bf
  k}\rangle$), and/or from competing  states such as a ZR
triplet, and/or singlets or triplets with the hole occupying a linear
combination of O orbitals with $s, p_x$ or $p_y$ instead of
$x^2-y^2$ symmetry. The latter possibility can also be ruled out
because overlaps with those Bloch states are found to be small. The
largest such contribution is from the ZR triplet state, shown in
Fig. \ref{fig8}(a) by the dashed line and open squares for $n_m=3,r$
without and with local spin fluctuations, respectively. This overlap
is much smaller than with the ZRS singlet.

Another way to confirm this is displayed in Fig. \ref{fig8}(b), where
we compare $p_{\rm ZRS}$ to $p_0+p_1$, where $p_0$ is the probability
to find the hole without any magnons, and $p_1$ is the probability to
find one magnon adjacent to the hole, in the quasiparticle
eigenstate. Note that $p_0+p_1 <1$ even for the one-magnon variational
approximation because the hole can also be located away from the
magnon. As $n_m$ increases, $p_0+p_1$ decreases even more as
configurations with two or more magnons now also contribute to the
normalization.  These configurations with two or more magnons, and
those with one magnon not adjacent to the hole, have no overlap with
$|{\rm ZRS},{\bf k}\rangle$, explaining the decrease in the magnitude
of $p_{\rm ZRS}$. However, the ratio $p_{\rm ZRS}/(p_0+p_1) > 0.9$ in
the whole Brillouin zone, confirming that this part of the
wavefunction has a predominant ZRS-like nature. This is certainly the
case near the $({\pi\over 2},{\pi\over 2})$ point, where the overlap
is converged to 1. Interestingly,  at the antinodal points this ratio
decreases with increasing $n_m$, and here the overlap with the ZR
triplet is largest, see Fig. \ref{fig8}(a), suggesting that a ZRS
description is less accurate in this region.  

Thus, the zero- and one- (adjacent) magnon parts of the wavefunction
have significant overlap with the ZRS Bloch state. However, $p_{\rm
  ZRS}\sim 0.5$ is a rather small value, and it is not clear whether
the dressing with more magnons is consistent with this ZRS
picture or not. It is possible that the two- and three-magnon components
of the wavefunction have significant overlap with a ZRS+one magnon and
ZRS+two magnon configurations, but they could also have quite
different nature. It is not clear to us how to verify which is the
actual situation.

If these two- and three-magnon components have significant non-ZRS
character, however that is defined in this case, then clearly the
difference observed in the results from one- and three-band models
would be likely due to this non-ZRS nature.

If, on the other hand, one takes these results to support the idea
that a low-energy projection onto 
ZRS states
is  valid, then this is not the origin of the
discrepancy in the quasiparticle behavior. In this case, it must
follow that:

(iii)  The $t$-$t'$-$t''$-$J$ is not the correct one-band model
because there are additional important terms generated by the projection onto
the ZRS states, like those discussed in 
Ref. \cite{ALigia} or the three-site terms, which it neglects.

If (iii) is indeed the explanation for the different behavior of the
quasiparticles of the one- and three-band models, then in our opinion
this implies that the strategy of using one-band models to study
cuprates is unlikely to succeed. The main reason for this strategy, as
mentioned, is to make the Hilbert space as small as possible for
computational convenience. This, however, is only useful if the
Hamiltonian is also fairly simple.

In principle one could test additional terms that could be included
in one-band models by using methods like ours, to figure out which
insure  that the resulting behavior mirrors that of
the three-band model. Even if this enterprise was successful and the ``fix'' was
relatively simple, {\em i.e.} only a few additional terms and
corresponding parameters are necessary, it is important to emphasize
that this improved one-band model would still {\em not} describe
correctly cuprates at finite doping. Additional terms must be included to
correctly account for the effective interactions between quasiparticles in
one-band models, as we demonstrated in Ref. \cite{Mirko}.
 
From a technical point of view, their origin is simple to understand.
Even for the simplified three-band model, the presence of additional
holes leads to additional terms in the Hamiltonian
\cite{BayoPRB}, because the intermediary states are different and this
affects the projection onto states with no-double occupancy on
Cu. This is going to become even more of an issue if a subsequent
projection onto ZRS-like states  has to be performed, and may well result in an
unmanageably complex Hamiltonian.

This is why we believe that the (simplified) three-band model is a
safer option to pursue. Its computational complexity is not that much
worse than for  one-band models, whereas the Hamiltonian is certainly
simpler. In fact, our demonstration here that there is no need to
accurately capture the spin fluctuations of the AFM background in
order to gain a reasonable understanding of the quasiparticle
behavior, makes its study significantly simpler. In particular, it
allowed us to study one hole on an infinite layer very simply and
efficiently. Generalizations to few holes \cite{MonaPRL} and to finite
concentrations could also turn out to be easier to carry out than the
effort of finding the correct form for a one-band Hamiltonian.

Of course, there is no guarantee that the (simplified) three-band
model captures all physics needed to explain cuprates, either. It is
possible that important aspects of the Emery model were lost through
the projection onto spin degrees of freedom at the Cu sites (this,
however, would affect the one-band models just as much). Even the
Emery model itself may not be general enough; for instance, a
generalization to a 5-band model including non-ligand O $2p$ orbitals
might be needed, as suggested recently in Ref. \cite{Hirsch}. We note
that such a generalization can be easily handled by our method
(provided that one can still project onto spin degrees of freedom at
the Cu sites), as showed in Ref. \cite{Hadi} where we found that these
states do not change the quasiparticle dispersion much, although they
do have an effect on its wavefunction.

While careful investigation of such scenarios is left as future work,
one clear lesson from this study is that obtaining the correct
dispersion for the quasiparticle of an effective model is {\em not
  sufficient} to validate that model. The dispersion can have the
correct shape for the wrong reasons, as we showed to be the case for
the $t$-$t'$-$t''$-$J$ model, where it is due to  the interplay between the
effects of the longer-range
hopping and spin-fluctuations. The same dispersion is obtained for
the simplified three-band model, however in this case the
spin-fluctuations play essentially no role, so the underlying physics
is very different. This difference is very likely to
manifest itself in other properties, therefore these models
are not equivalent despite the similar dispersion of their quasiparticles.

\acknowledgments
 We thank Walter Metzner and Peter Horsch for discussions
and suggestions. This work was supported by NSERC, QMI, and UBC 4YF (H.E.).


\begin{thebibliography}{99}
\bibitem{comment} Double occupancy is forbidden on Cu so here doping
  holes residing on O  interact with spins located on Cu.
\end{thebibliography}
\end{document}